\RequirePackage{amsmath}
\RequirePackage{amsfonts}

\documentclass{article}

\usepackage[T1]{fontenc}
\usepackage{graphicx}
\usepackage{hyperref}
\usepackage{color}
\usepackage{tabularx}
\usepackage{tikz}
\usetikzlibrary{fit,backgrounds}
\usepackage{authblk}

\begin{document}

\title{Embedding-based Methods for Linear Solver Performance Prediction}

\author{Hayden Liu Weng}
\author{Hans-Joachim Bungartz}
\author{Felix Dietrich}

\affil{Chair of Scientific Computing in Computer Science, Department of Computer Science, School of CIT, Technical University of Munich,\\
Boltzmannstr. 3, 85748 Garching, Germany;\\
h.liu@tum.de\\
\url{http://cs.cit.tum.de/sccs/}}

\date{}

\maketitle

\begin{abstract}
The solution of large, sparse linear systems often dominates the computational effort of scientific applications and is a frequent optimization target.
Modern libraries provide numerous solver and preconditioner configurations,
but their performance varies significantly across problem instances.
Previous works have addressed the selection of an optimal solver, but are typically limited by
the problem set addressed (e.g., only symmetric positive definite matrices), the use of expensive matrix features, or the complexity of the approach.

This work proposes a modular, low-cost embedding-based framework for solver selection that decouples performance modeling from feature representation and downstream prediction.
Solver-problem relationships are learned directly from observed performance data,
while inexpensive numerical features are used to project unseen problems into the learned embedding space.
The framework focuses on multilabel prediction and evaluation using user-centric metrics, such as MAPE and nDCG, which better reflect relative performance.

Experiments on 621 matrices from the SuiteSparse matrix collection across 101 PETSc solver configurations
demonstrate a 17\% increase in top-prediction accuracy over classical feature-based models
when expensive numerical features are included,
along with reductions of 37\% in mean average percentage error (MAPE) and 46\% in top-prediction error (1-error).
When restricted to a reduced feature set, the embedding approach remains competitive,
while still consistently achieving ca. 24\% lower MAPE and 1-error across a broad range of problems.
\end{abstract}

\section{Introduction}
In many scientific computing applications,
especially those stemming from the solution of differential equations,
one of the key steps involves solving an algebraic system of (possibly linearized) equations.
More importantly, this solution tends to be one of the main performance bottlenecks of the application.
Computing the solution to such systems is an ever-present challenge
due to numerical precision and stability concerns.
Since the resulting systems tend to be very large,
direct solvers are not ideal due to the memory requirements
and high computational complexity.
Consequently, the standard solution is the application of iterative methods
which, however, similarly suffer from limitations regarding convergence and numerical stability.

In order to address this, multiple algorithms exist, implemented in a variety of libraries,
but there is no silver bullet or universally best method.
Instead, each configuration can perform best in specific situations.
Not only can the optimal choice for a system be problem-dependent \cite{rice1976algorithm},
but the hardware itself may play a significant role,
particularly with modern trends towards a high degree of parallelization or the existence of accelerators (e.g., GPUs) \cite{xiao2023survey}.
Since expert knowledge may not always be readily available,
data-driven strategies have been long studied to tackle the solver selection problem.
However, these schemes typically rely on expensive features, including spectral information,
expensive models which might not scale well,
or are hard to generalize to other scenarios.

This work addresses whether learned embeddings,
i.e., low-dimensional representations of problems in which geometric proximity reflects similarity across solvers,
can replace these expensive matrix features while keeping overall model complexity low,
providing reliable solver predictions across diverse problem scenarios.
Moreover, the evaluation presented focuses on both ranking performance and the mean relative performance of the individual predictions,
better assessing the actual effectiveness of predictions in contrast to standard accuracy metrics.
In contrast to prior embedding-based approaches,
the proposed framework explicitly separates the performance-space embedding learning from feature-based projection of unseen matrices and downstream ranking-based prediction.
This modularization enables replacing individual components without retraining the full pipeline
and allows evaluation using multilabel ranking metrics aligned with relative performance objectives.

\section{Background} \label{sec:BG}

\subsection{Solving Preconditioned Linear Systems}

Solving a linear system with numerical software requires the selection of an appropriate solution algorithm and its implementation.
Since many problems of interest stem from very large systems,
direct methods tend to be out of the question due to their large memory requirements,
likely beyond what the user has available. 
Even when feasible, direct methods may be less efficient in practice,
particularly with lower tolerance requirements.
As a result, in the remainder of the paper, we focus only on iterative schemes.

In modern libraries and software, users can typically call a generic solution routine,
which tries to figure out a suitable solver scheme.
Typically, the software might check whether the problem is symmetric positive definite,
in which case the Conjugate Gradient (CG) method is the standard choice \cite{barrett1994templates,saad_iterative_book}.
Otherwise, other options might be applied for general symmetric matrices (e.g., MinRes) and for unsymmetric cases (e.g., QMR, BiCGStab, or GMRES) respectively.
Occasionally, a ``fallback'' method might be applied. 
While such heuristics provide a reasonable starting point, they fail to address preconditioning,
which has been widely shown to strongly affect solver performance \cite{benzi2002preconditioning,ferronato2012preconditioning,pearson2020preconditioners}.
Indeed, a solver that performs poorly on its own may still yield a good result with appropriate preconditioning.

\subsection{PETSc}

The Portable, Extensible Toolkit for Scientific Computing \cite{petsc-web-page} library is a commonly used framework
providing scalable solvers for linear and nonlinear problems and serves as the experimental platform for this study.
PETSc further facilitates experimentation by enabling compatibility with external libraries such as Hypre \cite{hypre}.
As a default setting, PETSc applies the restarted GMRES solver,
preconditioned with ILU(0) in the single-process case
or block Jacobi in the multi-process case.
These defaults provide reasonable robustness but are rarely optimal across diverse problem classes.
For this study, we consider 101 solver configurations spanning Krylov methods (CG, GMRES, BiCGStab, TFQMR, etc.) and algebraic multigrid, incomplete factorization, and approximate inverse preconditioners (e.g., BoomerAMG, ILU, ParaSails).

\subsection{Measuring Performance}

Many studies predicting solver performance focus primarily on runtime, $t$, as the main indicator of solver effectiveness.
However, even under the same settings, there may be schemes leading to very fast but inaccurate solutions in terms of residual value at convergence,
due to the effect of the preconditioner.
Therefore, fast solutions may still be too inaccurate for practical purposes.
As a result, and in order to incorporate respective time and accuracy requirements, we also consider the true (i.e., unpreconditioned) residual, $r$, of the solution upon convergence.
Both of these can then be combined into a score $s$, where

\begin{equation}
s\left(t,r\right)=\left(1+\log\left(1+\frac{w_t}{t}\right)\right)\left(1+\log\left(1+\frac{w_r}{r}\right)\right)-1.
    \label{eq:score}
\end{equation}
Here $w_t$ and $w_r$ correspond to predefined weights (by default $w_t=w_r=1$) indicating the importance of the respective criterion.
By log-transforming the input and shifting by one,
it is possible to naturally focus on individual performance criteria by setting the other weight to 0.
It is worth noting that lower runtime and residual values correspond to higher scores under this definition.
If a problem does not converge, then the score is set to 0.

\subsection{The Classification Problem}

When predicting solver performance,
directly predicting the runtime or residual norms introduces unnecessary complexity and overhead.
Instead, most existing approaches treat solver selection as a binary classification task,
separating effective solvers from ineffective ones;
or as a multiclass classification, directly predicting a single solver recommendation.
However, both of these approaches treat all misclassifications identically,
regardless of proximity to the optimal solver:
For instance, mislabeling the best solver in a binary setting, or predicting the second-best solver in a multiclass problem.

In practice, users are often interested in finding a suitable, ``good enough'' combination, even if it might not be the absolute best,
since the cost of further optimization could overshadow potential improvement.
We instead formulate the solver selection problem as a binary multilabel classification problem.
For each matrix, the model predicts a subset of promising solvers.
This choice additionally allows comparisons to previous approaches,
as the labels can still be evaluated independently,
and the solver with the highest prediction probability can be taken as the single top recommendation.
Moreover, this formulation naturally induces a ranking based on the internal probability values,
allowing the use of metrics relating to the relative ordering.

\subsubsection{Binary and multiclass metrics}

For comparability with classical approaches, we report \textit{accuracy} of the top-predicted class
and the \textit{one-error}, measuring how often the optimal solver is absent from the predicted set.
Since numerical performance values are available, we also compute the relative error between the top predicted solver and the best solver per instance via the \textit{Mean Average Percentage Error} (MAPE):

\begin{equation}
    \text{MAPE}(y_\text{true},\hat{y})=\frac{1}{n_\text{samples}}\sum_{i=0}^{n_\text{samples}-1}\frac{|s(y_{\text{true},i})-s(\hat{y}_i)|}{s(y_{\text{true},i})}\;,
    \label{eq:MAPE}
\end{equation}
for the top predicted solvers $\hat{y}$, and oracle solvers $y_\text{true}$.
A value of 0 corresponds to perfect prediction, and 1 to failing/divergent solvers for all instances.

\subsubsection{Multilabel metrics}

When explicitly treating the problem as a multilabel classification task,
focus shifts to the ordered set of predicted classes and whether these indeed correspond to effective solvers (true labels).
We consider \textit{Label Ranking Average Precision} (LRAP) \cite{tsoumakas2010mining-lrap} and \textit{Normalized Discounted Cumulative Gain} (nDCG) \cite{jarvelin2002cumulated-ndcg1,wang2013theoretical-ndcg2,mcsherry2008computing-ndcg3} as our metrics of choice.

LRAP is defined by
\begin{equation}
    \text{LRAP}(\mathbf{y}_\text{true},\hat{p}) = \frac{1}{n_\text{samples}} \sum_{i=0}^{n_\text{samples}-1} \left( \frac{1}{\|\mathbf{y}_{\text{true},i}\|_{0}} \sum_{j: \mathbf{y}_{\text{true},ij} = 1} \frac{\left|\mathcal{L}_{ij}\right|}{\text{rank}_{ij}}\right),
    \label{eq:lrap}
\end{equation}
where $\mathbf{y}_\text{true}\in \{0,1\}^{n_\text{samples}\times n_\text{labels}}$ are the true labels,
$\hat{p} \in [0,1]^{n_\text{samples}\times n_\text{labels}}$ the prediction probabilities,
$ \mathcal{L}_{ij} = \{k: \mathbf{y}_{\text{true},ik} = 1, \hat{p}_{ik} \geq \hat{p}_{ij} \} $ describes the number of positive labels seen up to $j$,
and $ \text{rank}_{ij} = |\{k: \hat{p}_{ik} \geq \hat{p}_{ij} \}| $ corresponds to the number of labels ranked higher or equal to $j$.
Since only the ranking itself is considered, mislabeled entries are penalized based on the proportion of correct and incorrect predictions.

The NDCG is derived from the DCG score computed for the whole vector:

\begin{equation}
    \text{DCG}\left(\mathbf{y}_\text{true},\hat{p}\right) = \sum_{r=1}^{n_\text{labels}} \frac{\mathbf{y}_{\hat\sigma(i)}}{\log\left(1+i\right)}\;,
    \label{eq:ndcg}
\end{equation}
with $\hat{\sigma(r)}$ the induced ranking by the prediction. Here, $\mathbf{y}_{\hat\sigma(i)}$ measures the relevance of the rank $r$ label (1 for true or 0 for false in our case).
The NDCG score is then defined as the DCG divided by the DCG score of a perfect ranking.
In contrast to the previous metric, the NDCG score now additionally penalizes misranked (missed true configurations) by the position of the error, rather than just by the proportion of values seen thus far.

\section{Related Work}  \label{sec:RelWork}

Achieving high performance requires tailoring algorithmic choices to the specific characteristics of the problem and computational context,
often resulting in solutions that are not easily transferable to other scenarios.
Initiatives such as the \textit{Self-Adapting Numerical Software (SANS-Effort)} \cite{sans,sansIEEE} have sought to address this challenge by automating the adaptation process.
Notable libraries following this philosophy include ATLAS \cite{whaley1998automatically} and LAPACK implementations in HPC \cite{sansLAPACK}, focusing on optimized configurations for specific hardware and problem domains.

For iterative linear solvers specifically, several studies have demonstrated the efficacy of traditional machine learning models predicting performance based on problem features \cite{bhowmick2006ml,george2008recommendation,jessup2016performance}. 
However, the commonly used metrics do not fully capture the practical impact of solver mispredictions,
where recommending a less-optimal solver may be acceptable, while recommending a diverging solver is catastrophic.
Beyond the traditional models, Yeom et al.~\cite{yeom2016data} introduce a performance vector space and nearest-neighbor classification to guide solver selection,
providing a demonstration of embedding-based approaches in solver selection.
They additionally introduce the use of user-centric performance measures such as MAPE,
which precisely allows differential assessment of mispredictions.
However, their analysis remains at the single-label recommendation level,
and mostly centers on SPD problems,
with results suggesting that expensive features were necessary to obtain more accurate predictions.
In contrast, the present work formulates the solver selection as a multilevel problem and explicitly decouples the pipeline's individual components.

Other contemporary techniques, such as deep learning approaches \cite{funk2022prediction},
graph neural networks (GNNs) \cite{tang2022gnn},
and convolutional neural networks (CNNs) \cite{xiong2025mm-autosolver},
have further expanded the portfolio of applicable techniques.
While promising, these approaches introduce substantial training and per-instance inference overheads
and may struggle to generalize across heterogeneous solver configurations, problem classes, and hardware environments.
Overall, while substantial progress has been made in both algorithmic development and data-driven selection techniques,
there remains a need for frameworks that explicitly account for practical performance trade-offs,
remain accessible to non-expert users,
and support different predictive mechanisms, problem types, and evaluation criteria in a unified manner.

\section{Embedding-based Prediction} \label{sec:Method}

\begin{figure}
\centering
\begin{tikzpicture}[
    every node/.style={
        rectangle,
        rounded corners,
        draw=black,
        align=center,
        font=\small
    },
    x=1.6cm,   
    y=1.0cm,    
    scale=0.85,
    transform shape
]

\def\ytop{1.8}
\def\ymtop{0.9}
\def\ymid{0}
\def\ymbot{-0.9}
\def\ybot{-1.8}

\def\xleft{-2.25}
\def\xtrain{0}
\def\xmodel{2.25}
\def\xinf{4.5}
\def\xout{6.75}


\node (feat)  at (\xleft,\ytop)   [circle] {Feature\\Set, $f$};
\node (prob)  at (\xleft,\ymid)   [circle] {Problem\\Set, $\mathcal{P}_\text{tr}$};
\node (sol)   at (\xleft,\ybot)   [circle] {Solver\\Set, $\mathcal{A}$};

\node (score) at (\xtrain,\ybot) [circle]
  {Score\\Metric\\$\ell$};

\node (extr)  at (\xtrain,\ytop)  {Feature\\Extraction\\$f(\mathcal{P}_\text{tr})$};
\node (coll)  at (\xtrain,\ymid)  {Data\\Collection\\$\ell(\mathcal{A},\mathcal{P}_\text{tr})$};

\node (sel)   at (\xmodel,\ymtop)  {Feature\\Selection\\$\hat{f}(\mathcal{P}_\text{tr})=:\mathcal{F}$};

\node (emb)   at (\xmodel,\ymbot)
  {Embedding\\
   $\{ g(x)\mid x\in\mathcal{P}_\text{tr}\}$\\
   $=:\mathcal{G}$};


\node (inst) at (\xinf,\ytop) [circle]
  {New\\Problem\\$x'\in\mathcal{P}$};

\node (proj) at (\xinf,\ymid)
  {Projection\\
   $\alpha:\,\hat{f}(x')\approx\alpha\cdot\mathcal{F}$\\
   $\Rightarrow \hat{g}(x'):=\alpha\cdot\mathcal{G}$};

\node (pred) at (\xinf,\ybot)
  {Prediction\\$\approx\ell(\mathcal{A},x')$};


\draw[->] (prob) -- (coll);
\draw[->] (sol) -- (coll);
\draw[->] (score) -- (coll);

\draw[<-] (extr) -- (prob);
\draw[->] (feat) -- (extr);

\draw[<-] (sel) -- (extr);
\draw[<-] (sel) -- (coll);
\draw[<-] (emb) -- (coll);

\draw[<-] (proj) -- (sel);
\draw[<-] (proj) -- (emb);
\draw[->] (inst) -- (proj);

\draw[<-] (pred) -- (proj);


\begin{scope}[on background layer]

\node[draw,
      dashed,
      rounded corners,
      fill=gray!6,
      inner sep=8pt,
      fit=(prob) (sol) (coll) (extr) (feat) (sel) (emb) (score)
     ] {};

\end{scope}
\end{tikzpicture}
\caption{Proposed pipeline, where $\mathcal{F}$ represents the collected values of selected features, $\mathcal{G}$ the embeddings for the training samples, and $\alpha$ the coefficients for the projection operator. Offline training phase indicated by the marked region.}
\label{fig:pipeline}
\end{figure}
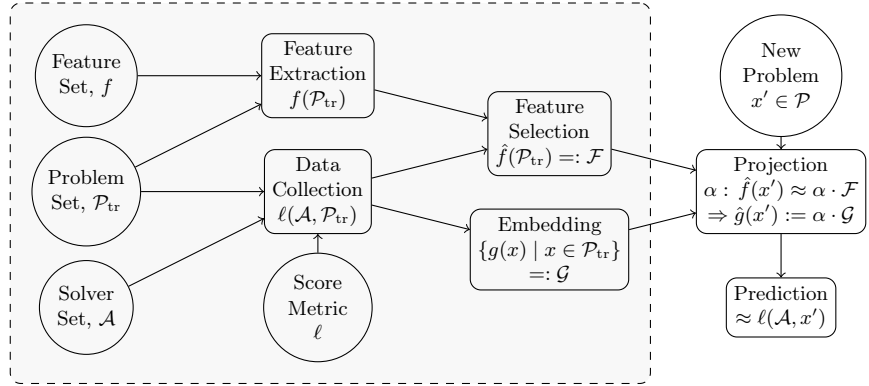

Instead of directly feeding property data and labels to a single traditional model,
we adopt a modular, embedding-based workflow for solver recommendation,
inspired by the performance vector space approach of Yeom et al. \cite{yeom2016data}.
That is, we learn a low-dimensional representation for the sample matrices,
where proximity in space reflects similarity in observed performance.
Unlike direct classification on matrix features,
the embedding captures additional relationships between matrices and solver configurations,
encoding solver similarity patterns that are not explicitly represented in the numerical descriptors.
Crucially, we extend the method introduced by Yeom et al. \cite{yeom2016data} and view the problem from a multilabel classification perspective.
As a result, each component in the framework can be easily exchanged without recomputing the rest of the pipeline, and the overall prediction quality is assessed using several metrics.
Figure~\ref{fig:pipeline} presents the complete pipeline of the method.

To define the problem space, a set of numerical features ($f$), training problems ($\mathcal{P}_\text{tr}$), solver configurations ($\mathcal{A}$), and metrics ($\ell$) are established.
These are then used for feature extraction, $f(\mathcal{P}_\text{tr})$, and performance data collection and labeling, $\ell(\mathcal{A},\mathcal{P}_\text{tr})$.
With the parsed input data, the pipeline then takes two steps, possibly in parallel.
Since the initial feature set can also contain more expensive or non-informative numerical properties,
it is possible to obtain a reduced feature set $\hat{f}\subseteq f$ by keeping only relevant features via feature selection.
In parallel, the performance data is used to learn embeddings $g(x)\in\mathcal{G}$ for the problems $x\in\mathcal{P}_\text{tr}$ in the training set,
encoding relative solver performance relationships across problems.
That is, the closer two problems are in the performance vector space, the more similar the solver behavior should be.
This way, the embedding encapsulates further relationships beyond the individual numerical features or their statistics.
Crucially, this separation allows different feature sets, embedding models, and prediction schemes to be exchanged independently, without altering the overall structure.

After the offline training phase of the pipeline,
once a new problem $x'$ comes in, the reduced feature set is computed for the new instance, and the two tracks are combined in order to obtain their projection $\hat{g}(x')$ onto the embedding space.
With this projection, a downstream predictor (e.g., k-nearest neighbors) is used to infer solver suitability.
Unlike prior embedding-based approaches that ultimately reduce solver selection to a single-label decision,
the proposed pipeline also supports ranking- and set-based predictions aligned with relative performance metrics.

\subsection{Feature Selection}

As the numerical features collected tend to comprise a broader set of entries,
the first step of the framework corresponds to feature selection,
also using the performance data.
After standard filtering out features with almost zero variance or too high correlations,
we apply the Boruta ``all-relevant'' feature selection scheme \cite{boruta-kursa2010feature}.
In contrast to classic selection schemes that aggressively optimize for the training data to provide minimal subsets of features,
Boruta seeks to keep all relevant features, i.e.,
those that perform better than their \textit{shadow} counterparts,
i.e., randomly shuffled versions of the data.
As a result, the determined set is less sensitive to the specific data provided.

In addition, costly features are excluded from the final set of features, such that the end result consists only of features that can be computed in $O(N)$ or $O(nnz)$ time, with $N$ being the number of rows and $nnz$ the number of nonzeros. Table~\ref{tab:features} illustrates the set of features considered, with selected features marked, and the final reduced set in bold.
In particular, this means that no direct estimate for the condition number is included in the final set.

\begin{table}
\caption{Matrix feature overview with properties as defined by Sood \cite{sood2019iterative}. Reduced set marked in \textbf{bold}} \label{tab:features}
\centering
\small
\setlength{\tabcolsep}{3pt}
\renewcommand{\arraystretch}{0.95}
\begin{tabularx}{\linewidth}{>{\centering\arraybackslash}X 
                                  >{\centering\arraybackslash}X
                                  >{\centering\arraybackslash}X
                                  >{\centering\arraybackslash}X}
\hline
\textbf{NumRows}* & NumCols & RowVariance & ColVariance \\
\textbf{DiagVariance}* & \textbf{Nonzeros}* & FrobeniusNorm & SymFrobeniusNorm \\
AntiSymFrobNorm & OneNorm & InftyNorm & SymInftyNorm \\
AntiSymInftyNorm & Trace & AbsTrace & \textbf{MaxNNZperRow}* \\
MinNNZperRow & AvgNNZperRow & DummyRows & DummyRowsKind \\
BoolSymmetry & BoolPatternSym & ValueSymFrac & PatternSymFrac \\
RowDiagDom & ColDiagDom & DiagAverage & DiagSign \\
\textbf{DiagNNZ}* & \textbf{LowerBW}* & \textbf{UpperBW}* & \textbf{RowLogValSpread}* \\
\textbf{ColLogValSpread}* & GerschgorinMax & \textbf{GerschgorinMin}* & KappaEstimate* \\
\hline
\end{tabularx}
$^*$ Features selected by Boruta, KappaEstimate is excluded due to computational cost
\end{table}

\subsection{Embedding}

As indicated previously, the embedding is computed using only the performance data,
which corresponds to the \textit{performance vector space} from Yeom et al. \cite{yeom2016data}.
This separation is also useful later for mapping unseen data using the still-unused feature information.
As a result, matrices and solver configurations are independently treated as tokens,
to obtain a vocabulary with $N_\text{matrices}+N_\text{solvers}$ elements.

The embedding thus learns mappings $g:\,\mathcal{P}\to\mathcal{G}\subseteq{\mathbb{R}}^d$
such that effective matrix–solver pairs are mapped to vectors close to each other, while ineffective pairs are separated.
As a consequence, solvers that tend to be effective for the same matrices will be mapped closer,
as will matrices with similar properties.
In practice, this objective can be implemented using embedding models such as skip-gram with negative sampling (\textit{word2vec}) \cite{mikolov2013word2vec} or \textit{GloVe} \cite{pennington2014glove,carlson2025new}, among others.

Matrix-solver pairs for effective combinations are then fed as co-ocurrences, whereas other pairs are used for negative sampling.
The coefficients for each pair of values are therefore adjusted accordingly.
Once embeddings have been computed for the training data, the offline phase is done.

\subsection{Projection}

Since embeddings are learned only for training problems, which are treated atomically,
unseen matrices do not have a direct representation in the embedding space.
We therefore compute a sparse linear combination of training feature vectors that approximates the feature vector of the new problem:

\begin{equation}
    \min_\alpha:\,\|\hat{f}(x')-\alpha\cdot\mathcal{F}\| \Rightarrow  \hat{g}(x'):=\alpha\cdot\mathcal{G}\,.
    \label{eq:projection}
\end{equation}
For better interpretability of the resulting combination,
the model is constrained to non-negative coefficients only. 
Moreover, given the limited size of the feature set,
standard linear regression can achieve a good degree of sparsity without the need for additional penalization (e.g., via LASSO or Ridge regression).
The same coefficients can then be applied to the corresponding embeddings,
leading to the embedding for the unseen matrix.
Therefore, assuming consistency between the feature and embedding spaces,
the same coefficients are used to map the new sample naturally into the embedding space.

\subsection{Prediction}
Once the projection has been computed, it remains to predict the performance of the novel problem across the different solver configurations.
The downstream predictor directly operates in the learned embedding space and produces a multilabel output,
remaining decoupled from the previous elements of the framework.
For this work, a simple k-nearest neighbors algorithm was used.

As a result, at inference time, the proposed framework requires only the computation of the reduced feature set and projection onto the learned embedding space,
which is considerably less work than other state-of-the-art approaches.

\section{Experimental Setup} \label{sec:Setup}

\subsection{Hardware and Software Environment}

Performance measurements were conducted on a cluster node to ensure realistic large-scale parallel behavior,
while feature computation, model training, and prediction were performed on a separate workstation.
This emulates many common workflows using a separate machine for development.
The performance data was collected on a single node of the LRZ CoolMUC-4 Cluster\footnote{\url{https://doku.lrz.de/linux-cluster-10745672.html}},
equipped with dual Intel Xeon Platinum 8480+ (Sapphire Rapids) CPUs with a total of 112 processors and 512 GB of RAM.
Each problem in the dataset was solved in parallel using 112 MPI processes using PETSc 3.22.1 compiled with GCC 13.2.0 and openMPI 5.0.8.
The remainder of the operations, including both gathering the property data and training the model,
were instead computed on a workstation equipped with an 8-core Intel Core i7-10700 (Comet Lake) CPU,
32 GB of RAM, and an Nvidia Quadro RTX 5000 GPU with 16GB of memory.
The prediction framework used scikit-learn 1.6.1, BorutaPy 0.4.3, and gensim 4.4.0.

\subsection{Dataset Construction and Solver Configuration}

The matrices used for training were taken from the SuiteSparse matrix collection \cite{SuiteSparseCollection}.
Only square matrices with real values and dimensions between 1000 and 10000 rows
and $\leq$ 200'000 nonzeros were considered for broader solver coverage, as rectangular (least-squares) problems require different approaches.
As no further filters were added, this resulted in 621 matrices, including symmetric positive definite, general symmetric, and unsymmetric problems.
The right-hand side was fixed to the vector of ones, and an initial guess of 0 was used throughout all tests.

As for solver settings in PETSc, these were kept at their default values of \texttt{rtol=1e-5}, \texttt{atol=1e-50}, and \texttt{dtol=1e5}, although the maximum number of iterations was increased to \texttt{maxits=1e5} to allow additional iterations to compensate for potential numerical inaccuracies.
A small dummy system was computed in advance of each timed run to preload instructions and reduce cold-cache effects.
Additionally, a maximum runtime of 600~s was set for each solve,
to allow some of the more expensive but accurate solver configurations to complete.
As a result, the number of instances that timed out was rather small.

\subsection{Model Training and Baselines}

The collected data was also used to train other classic machine learning methods, including k-nearest neighbors and random forests.
Each of these additional models was trained with the full set of properties,
to determine whether the embedding using the reduced feature set performed competitively with these other models.

For all trained models, we consider two scenarios: time-only score and time-residual score.
Solver configurations were deemed effective for a problem instance if their performance fell within 10\% of the top score.
While alternative thresholds are possible, sensitivity analysis is left for future work.
All metrics were computed under 5-fold cross-validation,
with suitable score metrics also being computed against the single best configuration across all samples, denoted as the ``single'' strategy.

\section{Results and Discussion} \label{sec:Results}

\begin{figure}
\includegraphics[width=\textwidth,
]{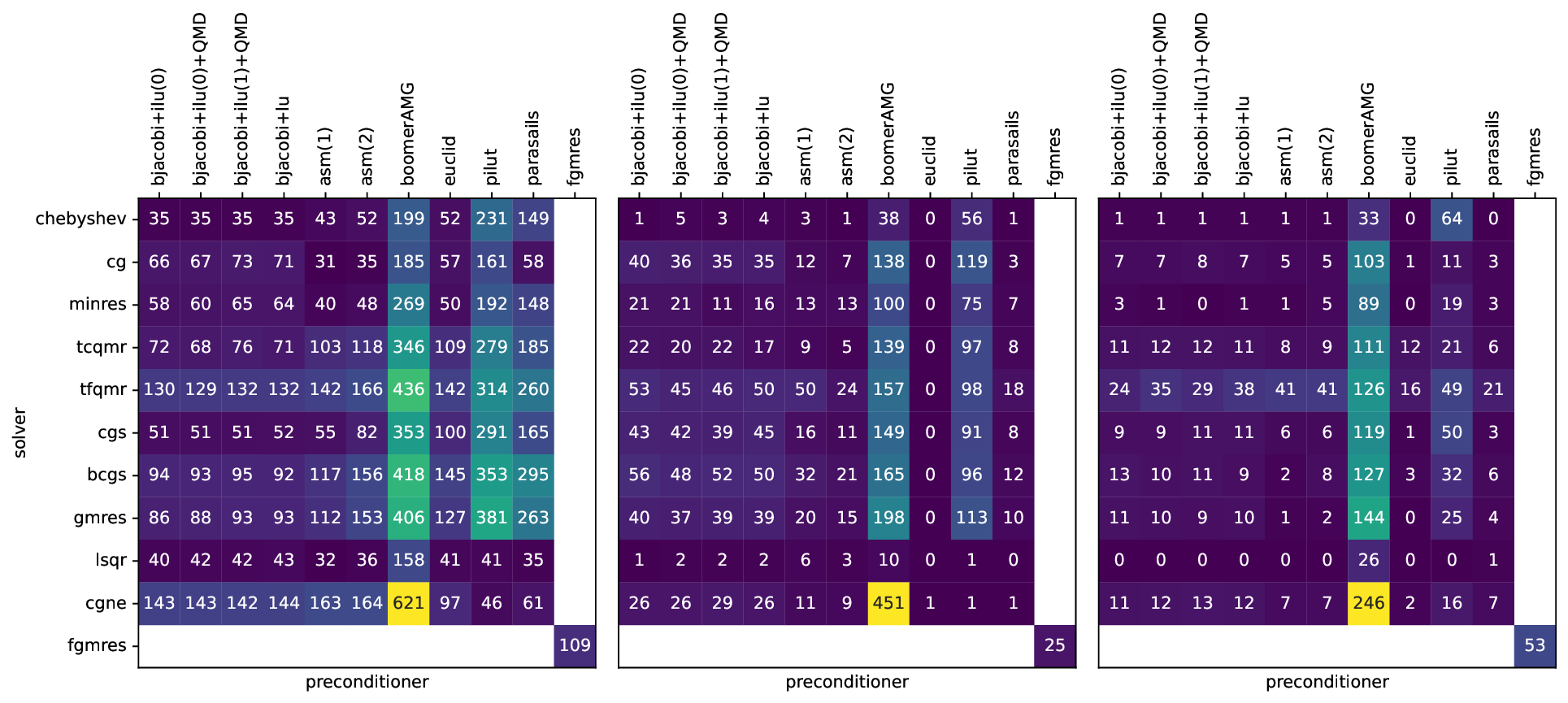}
\caption{Number of problems successfully converging (left), within 10\% of the best solver's time (center), and within 10\% of the best solver's score (right).} \label{fig:heatmaps}
\end{figure}

Figure~\ref{fig:heatmaps} provides a first impression of the performance data collected,
respectively showing the number of problems which converged, were effective with respect to time only, or yielded scores within 10\% of the maximum score on each problem, for each solver configuration.
The single top choice of CGNE~+~BoomerAMG converged for all matrices, possibly due to the amount of memory available to the system used.
Taking time into account, the combination remains effective for 72.6\% of cases, whereas with both time and residual, this drops to 39.6\%.
This top combination, however, still underperformed considerably in specific cases,
with performances up to 61.7\% and 97.9\% worse with respect to time-only and time-residual scores, respectively.
This coincides with the expectation that the normal equations form an SPD system, more easily leading to convergence,
at the cost of reduced numerical stability.

Other general trends can be identified, with the effect of preconditioners tending to be more noticeable across the data.
Solvers for non-symmetric problems seem to work well on distinct problems, with TFQMR, BiCGStab, and GMRES being especially effective.
TFQMR additionally remained slightly more effective even across preconditioners, highlighting its performance when convergent.
Consistent with the literature, however, the biggest factor remained the preconditioner.
Overall, the BoomerAMG, PILUT, and ParaSails were the most effective, with BoomerAMG clearly standing out for general problems.
Parasails, in particular, was less effective in score metrics due to the overhead of computing the approximate inverse.
Notably, the solver parameters were not fine-tuned, so a dedicated study could still yield more varied results.

\begin{figure}
\includegraphics[width=\textwidth,
]{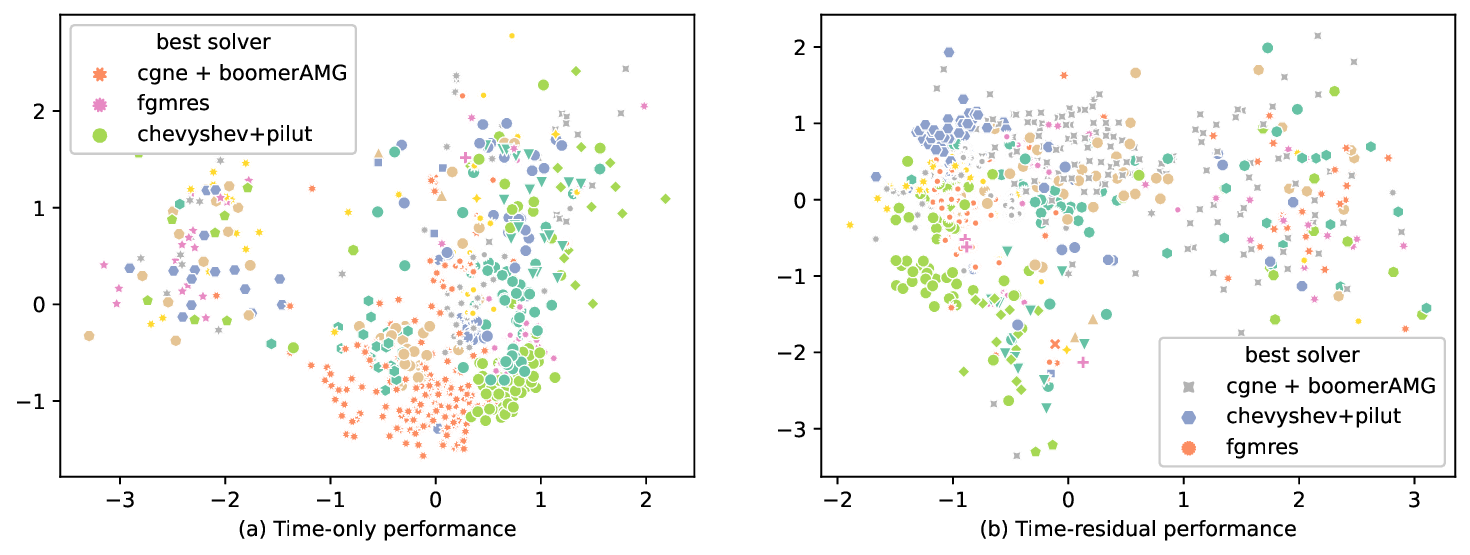}
\caption{PCA projections of the trained embeddings, colored by best solver. The three largest clusters are included in the legend.} \label{fig:embeddings}
\end{figure}

We next turn to the PCA of the learned embeddings in Figure~\ref{fig:embeddings},
where points are colored by their best solver.
The embedding effectively clusters samples with the same best solver,
while maintaining spread across the embedding space.
Rather than isolated clusters, we observe that clusters tend to still co-locate
since problems can have multiple effective solvers in addition to the top solver.
Samples with less common best solvers remain farther from the clusters due to their smaller support.

Table~\ref{tab:comparison-time-only} shows results when considering time-only scores, with most schemes showing similar accuracy and MAPE.
These results imply that all schemes successfully find a suitable solver in most instances.
Since the single-best strategy converged for all cases (leading to a non-zero performance score), its MAPE error is remarkably low.
In comparison, our methods still predicted a diverging solver for a few instances, thereby increasing the resulting error considerably.
Regardless, the reduced embedding remains competitive with classic approaches, achieving marginally higher nDCG and LRAP scores.
The largest difference is in the 1-error, as the best solver was not effectively recovered by classic methods in general.
In contrast, the proposed embedding significantly reduced the 1-error by 24\% with the reduced embedding and 41\% with the full feature set.

\begin{table}%
{\footnotesize
  \caption{Performance measurements across prediction schemes relative to time-only score.}  \label{tab:comparison-time-only}
\centering
    \begin{tabular}{|c||c|c|c|c|c|} \hline
        Method         & Accuracy      & MAPE          & 1-error & nDCG & LRAP \\
        \hline\hline
        Single         & 0.726$\pm$0.037 & 0.072$\pm$0.007 & 0.634$\pm$0.055
                        &  --- &  --- \\
        k-NN           & 0.708$\pm$0.046 & 0.147$\pm$0.013 & 0.639$\pm$0.050 
                        & 0.781$\pm$0.018 & 0.624$\pm$0.026 \\
        Random Forest  & 0.700$\pm$0.039 & 0.187$\pm$0.022 & 0.665$\pm$0.041 
                        & 0.787$\pm$0.016 & 0.649$\pm$0.022 \\ \hline
        Full Embedding & 0.786$\pm$0.046 & 0.125$\pm$0.016 & 0.270$\pm$0.036
                        & 0.827$\pm$0.015 & 0.707$\pm$0.016 \\
        Red. Embedding & 0.721$\pm$0.029 & 0.146$\pm$0.008 & 0.422$\pm$0.076
                        & 0.791$\pm$0.015 & 0.656$\pm$0.025 \\
        \hline
    \end{tabular}
}%
\end{table}%

Now focusing on the more strict time-residual scores, we observe a much lower number of effective combinations overall.
As a result, Table~\ref{tab:comparison-both} shows lower scores and higher errors across all models.
While the proposed method with the full feature set considerably outperforms the classical methods,
the reduced embedding performs at the same level as the Random Forest trained on the full feature set.
As before, the largest discrepancy can be seen in the error metrics:
The 1-error drops from 90.8\% for the Random Forest method and 76.5\% for the single strategy to 44.8\% and 58.6\% for the proposed method with full and reduced features, respectively.
Similarly, the MAPE error is significantly reduced to 24.8\% and 38.1\%, respectively.

\begin{table}%
{\footnotesize
  \caption{Performance measurements across prediction schemes relative to time-residual score}  \label{tab:comparison-both}
\centering
    \begin{tabular}{|c||c|c|c|c|c|} \hline
        Method         & Accuracy       & MAPE             & 1-error & nDCG & LRAP \\
        \hline\hline
        Single         & 0.396$\pm$0.050 & 0.417$\pm$0.035 & 0.765$\pm$0.021
                        &  ---            &  ---            \\
        k-NN           & 0.396$\pm$0.045 & 0.614$\pm$0.048 & 0.948$\pm$0.021
                        & 0.599$\pm$0.028 & 0.431$\pm$0.031 \\
        Random Forest  & 0.430$\pm$0.038 & 0.624$\pm$0.035 & 0.908$\pm$0.036
                        & 0.626$\pm$0.025 & 0.484$\pm$0.030 \\ \hline
        Full Embedding & 0.601$\pm$0.028 & 0.248$\pm$0.023 & 0.448$\pm$0.008
                        & 0.724$\pm$0.017 & 0.610$\pm$0.025 \\
        Red. Embedding & 0.420$\pm$0.025 & 0.381$\pm$0.028 & 0.586$\pm$0.029
                        & 0.606$\pm$0.019 & 0.472$\pm$0.022 \\
        \hline
    \end{tabular}
}%
\end{table}%

\begin{figure}%
\includegraphics[width=\textwidth,
]{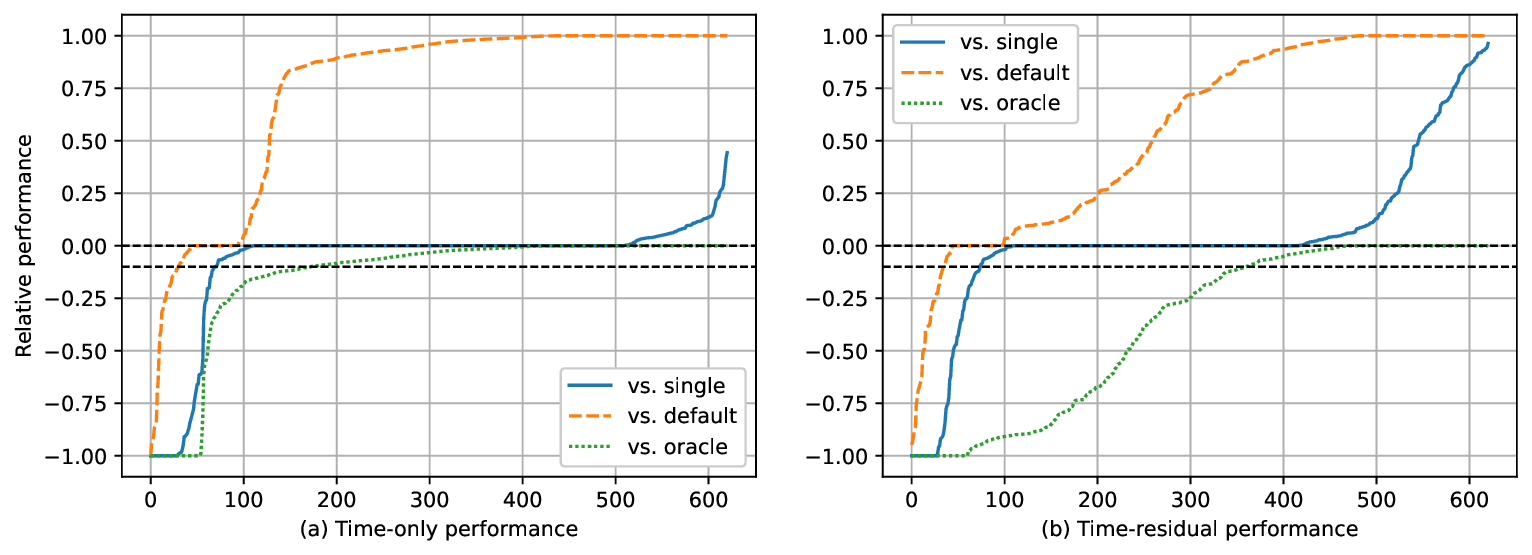}
\caption{Sorted relative performance of the reduced embedding predictor on individual samples across all matrices compared to other strategies.} \label{fig:rel-perf}%
\end{figure}

As a final evaluation, Figure~\ref{fig:rel-perf} displays relative errors for the predictions across the problem instances when compared against the oracle solver, the single-best solver, and the default PETSc solver.
For the time-only score, the proposed method quickly finds the best solver roughly 50\% of the time, while still returning an effective solver on 75\% of problems.
For the time-residual score, the panorama is less favorable: the best solver is predicted 25\% of the time, and an effective solver in 43\% of cases, with performance decaying gradually beyond that point.
In either case, the proposed method only fails to predict a convergent solver in less than 10\% of cases.
Nevertheless, the proposed method still performs better than the default PETSc configuration,
and at least as well as the single-best solver for over 80\% of tested cases.
Since applying the solver prediction scheme does not require expensive feature computations,
the resulting prediction should in general lead to beneficial results.

\section{Conclusions and Outlook}  \label{sec:Concl}
In this work, we presented a modular embedding-based framework for solver selection for sparse linear systems.
The approach comprised three main components: learned embeddings from performance data,
projection of unseen matrices using inexpensive numerical features,
and multilabel prediction with relative performance metrics.
Experiments on 621 matrices and 101 PETSc solver configurations show that the embedding-based model
consistently outperforms classical feature-based approaches,
while remaining competitive when restricted to low-cost features.
In particular, the reduced model achieves substantially lower MAPE and 1-error,
resulting in predictions that more reliably include near-optimal solvers.
The modular architecture also allows the reuse and extension of individual components without retraining the full pipeline.
By reducing reliance on expensive feature extraction,
the approach lowers inference overhead and offers a scalable alternative to direct feature-based classification.

The effectiveness of the embedding depends on sufficient diversity and coverage in the training data,
being less effective for solvers or matrices that are poorly represented.
A simple mitigation strategy could identify likely failures and revert to a fallback solver (e.g., the single-best method) instead.
Moreover, the projection step relies on a linear combination of the matrix features,
which may fail to capture nonlinear relationships across samples.
Hyperparameter tuning and systematic sensitivity analysis were similarly beyond the scope of this work,
but may further improve robustness and prediction performance.%

Future work includes exploring nonlinear projection mechanisms, incorporating richer structural representations, and extending the framework to hardware-aware prediction to improve portability across systems.
Ultimately, extending the approach to broader numerical tasks (e.g., least-squares problems, non-linear solvers) and general algorithm selection problems may further demonstrate the general applicability of the performance-space modeling approach for scientific computing workflows.

\subsection*{Acknowledgements} The authors gratefully acknowledge the computational and data resources provided by the Leibniz Supercomputing Centre (www.lrz.de).
F.D. acknowledges support by the DFG, project no. 468830823.

\subsection*{Disclosure of Interests}
The authors have no competing interests to declare that are relevant to the content of this article.

\appendix

\section{Data availability}

All input files, job scripts, models and code used to generate the figures can be found at \url{https://gitlab.com/h.liu/embedding-based-predictor}.

\bibliographystyle{plain}
\bibliography{references}

@incollection{rice1976algorithm,
  title={The algorithm selection problem},
  author={Rice, John R},
  booktitle={Advances in computers},
  volume={15},
  pages={65--118},
  year={1976},
  publisher={Elsevier}
}

@inproceedings{whaley1998automatically,
  title={Automatically tuned linear algebra software},
  author={Whaley, R Clinton and Dongarra, Jack J},
  booktitle={SC'98: Proceedings of the 1998 ACM/IEEE conference on Supercomputing},
  pages={38--38},
  year={1998},
  organization={IEEE}
}

@article{sans,
  title={Self-adapting numerical software for next generation applications},
  author={Dongarra, Jack and Eijkhout, Victor},
  journal={The International Journal of High Performance Computing Applications},
  volume={17},
  number={2},
  pages={125--131},
  year={2003},
  publisher={SAGE Publications}
}

@article{sansIEEE,
  title={Self-adapting numerical software (SANS) effort},
  author={Dongarra, Jack and Bosilca, George and Chen, Zizhong and Eijkhout, Victor and Fagg, Graham E and Fuentes, Erika and Langou, Julien and Luszczek, Piotr and Pjesivac-Grbovic, Jelena and Seymour, Keith and others},
  journal={IBM Journal of Research and Development},
  volume={50},
  number={2.3},
  pages={223--238},
  year={2006},
  publisher={IBM}
}

@article{sansLAPACK,
  title={Self-adapting software for numerical linear algebra and LAPACK for clusters},
  author={Chen, Zizhong and Dongarra, Jack and Luszczek, Piotr and Roche, Kenneth},
  journal={Parallel Computing},
  volume={29},
  number={11-12},
  pages={1723--1743},
  year={2003},
  publisher={Elsevier}
}

@book{barrett1994templates,
  title={Templates for the solution of linear systems: building blocks for iterative methods},
  author={Barrett, Richard and Berry, Michael and Chan, Tony F and Demmel, James and Donato, June and Dongarra, Jack and Eijkhout, Victor and Pozo, Roldan and Romine, Charles and Van der Vorst, Henk},
  year={1994},
  publisher={SIAM}
}

@book{saad_iterative_book,
  title={Iterative methods for sparse linear systems},
  author={Saad, Yousef},
  year={2003},
  publisher={SIAM}
}

@article{benzi2002preconditioning,
  title={Preconditioning techniques for large linear systems: a survey},
  author={Benzi, Michele},
  journal={Journal of computational Physics},
  volume={182},
  number={2},
  pages={418--477},
  year={2002},
  publisher={Elsevier}
}

@article{ferronato2012preconditioning,
  title={Preconditioning for sparse linear systems at the dawn of the 21st century: History, current developments, and future perspectives},
  author={Ferronato, Massimiliano},
  journal={International Scholarly Research Notices},
  volume={2012},
  number={1},
  pages={127647},
  year={2012},
  publisher={Wiley Online Library}
}

@article{pearson2020preconditioners,
  title={Preconditioners for Krylov subspace methods: An overview},
  author={Pearson, John W and Pestana, Jennifer},
  journal={GAMM-Mitteilungen},
  volume={43},
  number={4},
  pages={e202000015},
  year={2020},
  publisher={Wiley Online Library}
}

@article{xiao2023survey,
  title={A survey of accelerating parallel sparse linear algebra},
  author={Xiao, Guoqing and Yin, Chuanghui and Zhou, Tao and Li, Xueqi and Chen, Yuedan and Li, Kenli},
  journal={ACM Computing Surveys},
  volume={56},
  number={1},
  pages={1--38},
  year={2023},
  publisher={ACM New York, NY, USA}
}

@article{bhowmick2006ml,
  title={Application of machine learning to the selection of sparse linear solvers},
  author={Bhowmick, Sanjukta and Eijkhout, Victor and Freund, Yoav and Fuentes, Erika and Keyes, David},
  journal={Int. J. High Perf. Comput. Appl},
  year={2006},
  publisher={Citeseer}
}

@inproceedings{george2008recommendation,
  title={A recommendation system for preconditioned iterative solvers},
  author={George, Thomas and Gupta, Anshul and Sarin, Vivek},
  booktitle={2008 Eighth IEEE International Conference on Data Mining},
  pages={803--808},
  year={2008},
  organization={IEEE}
}

@article{jessup2016performance,
  title={Performance-based numerical solver selection in the Lighthouse framework},
  author={Jessup, Elizabeth and Motter, Pate and Norris, Boyana and Sood, Kanika},
  journal={SIAM Journal on Scientific Computing},
  volume={38},
  number={5},
  pages={S750--S771},
  year={2016},
  publisher={SIAM}
}

@phdthesis{sood2019iterative,
  title={Iterative solver selection techniques for sparse linear systems},
  author={Sood, Kanika},
  year={2019},
  school={University of Oregon}
}

@inproceedings{yeom2016data,
  title={Data-driven performance modeling of linear solvers for sparse matrices},
  author={Yeom, Jae-Seung and Thiagarajan, Jayaraman J and Bhatele, Abhinav and Bronevetsky, Greg and Kolev, Tzanio},
  booktitle={2016 7th International Workshop on Performance Modeling, Benchmarking and Simulation of High Performance Computer Systems (PMBS)},
  pages={32--42},
  year={2016},
  organization={IEEE}
}

@inproceedings{tang2022gnn,
  title={Graph Neural Networks for Selection of Preconditioners and Krylov Solvers},
  author={Tang, Ziyuan and Zhang, Hong and Chen, Jie},
  booktitle={NeurIPS 2022 Workshop: New Frontiers in Graph Learning},
  year={2022}
}

@inproceedings{funk2022prediction,
  title={Prediction of optimal solvers for sparse linear systems using deep learning},
  author={Funk, Yannick and G{\"o}tz, Markus and Anzt, Hartwig},
  booktitle={Proceedings of the 2022 SIAM Conference on Parallel Processing for Scientific Computing},
  pages={14--24},
  year={2022},
  organization={Society for Industrial and Applied Mathematics}
}

@article{xiong2025mm-autosolver,
  title={MM-AutoSolver: A multimodal machine learning method for the auto-selection of iterative solvers and preconditioners},
  author={Xiong, Hantao and Yang, Wangdong and He, Weiqing and Lin, Shengle and Li, Keqin and Li, Kenli},
  journal={Journal of Parallel and Distributed Computing},
  pages={105144},
  year={2025},
  publisher={Elsevier}
}

@article{SuiteSparseCollection,
  title={The University of Florida sparse matrix collection},
  author={Davis, Timothy A and Hu, Yifan},
  journal={ACM Transactions on Mathematical Software (TOMS)},
  volume={38},
  number={1},
  pages={1--25},
  year={2011},
  publisher={ACM New York, NY, USA}
}

@Misc{            petsc-web-page,
  author        = {Satish Balay and Shrirang Abhyankar and Mark~F. Adams and Steven Benson and Jed
                  Brown and Peter Brune and Kris Buschelman and Emil~M. Constantinescu and Lisandro
                  Dalcin and Alp Dener and Victor Eijkhout and Jacob Faibussowitsch and William~D.
                  Gropp and V\'{a}clav Hapla and Tobin Isaac and Pierre Jolivet and Dmitry Karpeev
                  and Dinesh Kaushik and Matthew~G. Knepley and Fande Kong and Scott Kruger and
                  Dave~A. May and Lois Curfman McInnes and Richard Tran Mills and Lawrence Mitchell
                  and Todd Munson and Jose~E. Roman and Karl Rupp and Patrick Sanan and Jason Sarich
                  and Barry~F. Smith and Stefano Zampini and Hong Zhang and Hong Zhang and Junchao
                  Zhang},
  title         = {{PETS}c {W}eb page},
  url           = {https://petsc.org/},
  howpublished  = {\url{https://petsc.org/}},
  year          = {2024}
}

@misc{hypre,
  key   =       {hypre},
  title =       {{\sl hypre}: High Performance Preconditioners},
  note =        {\url{https://llnl.gov/casc/hypre}, \url{https://github.com/hypre-space/hypre}}
  }

@article{mikolov2013word2vec,
  title={Distributed representations of words and phrases and their compositionality},
  author={Mikolov, Tomas and Sutskever, Ilya and Chen, Kai and Corrado, Greg S and Dean, Jeff},
  journal={Advances in neural information processing systems},
  volume={26},
  year={2013}
}

@inproceedings{pennington2014glove,
  title={Glove: Global vectors for word representation},
  author={Pennington, Jeffrey and Socher, Richard and Manning, Christopher D},
  booktitle={Proceedings of the 2014 conference on empirical methods in natural language processing (EMNLP)},
  pages={1532--1543},
  year={2014}
}

@article{carlson2025new,
  title={A new pair of gloves},
  author={Carlson, Riley and Bauer, John and Manning, Christopher D},
  journal={arXiv preprint arXiv:2507.18103},
  year={2025}
}

@article{tsoumakas2010mining-lrap,
  title={Mining multi-label data},
  author={Tsoumakas, Grigorios and Katakis, Ioannis and Vlahavas, Ioannis},
  journal={Data mining and knowledge discovery handbook},
  pages={667--685},
  year={2010},
  publisher={Springer}
}

@article{jarvelin2002cumulated-ndcg1,
  title={Cumulated gain-based evaluation of IR techniques},
  author={J{\"a}rvelin, Kalervo and Kek{\"a}l{\"a}inen, Jaana},
  journal={ACM Transactions on Information Systems (TOIS)},
  volume={20},
  number={4},
  pages={422--446},
  year={2002},
  publisher={ACM New York, NY, USA}
}

@inproceedings{wang2013theoretical-ndcg2,
  title={A theoretical analysis of NDCG type ranking measures},
  author={Wang, Yining and Wang, Liwei and Li, Yuanzhi and He, Di and Liu, Tie-Yan},
  booktitle={Conference on learning theory},
  pages={25--54},
  year={2013},
  organization={PMLR}
}

@inproceedings{mcsherry2008computing-ndcg3,
  title={Computing information retrieval performance measures efficiently in the presence of tied scores},
  author={McSherry, Frank and Najork, Marc},
  booktitle={European conference on information retrieval},
  pages={414--421},
  year={2008},
  organization={Springer}
}

@article{boruta-kursa2010feature,
  title={Feature selection with the Boruta package},
  author={Kursa, Miron B and Rudnicki, Witold R},
  journal={Journal of statistical software},
  volume={36},
  pages={1--13},
  year={2010}
}

\end{document}